\documentclass[prl,amsmath,amssymb,superscriptaddress,twocolumn]{revtex4}
\usepackage{graphicx}

\renewcommand{\Psi}{\varPsi}

\begin{document}

\title{Spacetime topology change and black
hole information}

\author{Stephen~D.~H.~Hsu} \email{hsu@duende.uoregon.edu}
\affiliation{Institute of Theoretical Science, University of Oregon,
Eugene, OR 97403}

\begin{abstract}
Topology change -- the creation of a disconnected baby universe --
due to black hole collapse may resolve the information loss
paradox. Evolution from an early time Cauchy surface to a final
surface which includes a slice of the disconnected region can be
unitary and consistent with conventional quantum mechanics. We discuss
the issue of cluster decomposition, showing that any violations
thereof are likely to be unobservably small. Topology change is
similar to the black hole remnant scenario and only requires
assumptions about the behavior of quantum gravity in Planckian
regimes. It does not require non-locality or any modification of
low-energy physics.
\end{abstract}

%\pacs{} 

%\date{today}

\maketitle

\section{Introduction: black hole information loss} 

What is the ultimate fate of something that falls into a black hole?
Is it crushed out of existence at a singularity, or does it end up
``somewhere else''? The answer to this question is central to the
black hole information loss paradox \cite{Hawking}. We will examine
the possibility that black hole formation leads to spacetime topology
change, and that matter that falls through the horizon ultimately
reaches some topologically disconnected component of the universe,
referred to here as a baby universe.  This scenario leads to a
resolution of the paradox without non-locality or modifications of low
energy physics.

There are numerous excellent reviews \cite{review1}--\cite{review3} of
the black hole information problem, so we give only a brief overview
here.  Figure 1 depicts the spacetime of an initially large, but
subsequently evaporating, black hole. The dashed line (Cauchy slice 1)
indicates a Cauchy surface on which the Schrodinger wavefunction and
its derivatives, describing the matter and gravitational fields, fully
specify future evolution (we assume the universe is in a pure quantum
state at early times). Some of the information (on the part of slice 1
behind the event horizon) will never reach future infinity $B$. In
particular, the Hawking radiation from the hole is spacelike separated
from the information behind the horizon, so there is no mechanism for
its escape which does not violate causality and locality. (Some
interesting mechanisms for such non-locality have been proposed in
string theory \cite{review3} and quantum gravity \cite{Gi}.) A
description of physics on slice 2 is necessarily a mixed state if we
are required to trace over the lost information that falls into the
hole. Dire physical consequences related to energy non-conservation
have been deduced for systems in which pure states evolve into mixed
states \cite{BPS}.

There are two main objections against topology change as a solution to
the information problem, which we list here and address later in this
paper.

\noindent {\bf Objection I}. An effective description of the evolution
of a pure state on past infinity (region $A$ in Figure 1) to a mixed
state on future infinity (region $B$ in Figure 2) must be sick,
perhaps exhibiting energy non-conservation, due to the arguments of
\cite{BPS}.

\noindent {\bf Objection II}. Processes which produce topologically
disconnected universes generally lead to violation of cluster
decomposition (locality) \cite{Su}.

Earlier work on topology change and its relation to black hole
information includes an unpublished preprint by Dyson \cite{Dyson},
papers by Strominger \cite{St} and Polchinski and Strominger
\cite{PS}, and work by Jacobson \cite{Ja,review2b} and Easson and
Brandenberger \cite{EB}. Also of interest is the work of Ashtekar and
Bojowald \cite{AB}, in which quantization of the classical singularity
region of a black hole allows evolution of a large spacetime. The
authors of \cite{Dyson}, \cite{St}, \cite{Ja}, \cite{EB} and \cite{AB}
all state that new universe creation might alleviate the black hole
information problem, although the specific objections I and II on
which we focus are not fully addressed. In \cite{PS} the scheme of
third quantization (originally developed for spacetime wormholes
\cite{wormholes}) is used, which leads to a peculiar kind of
indeterminacy. We do {\it not} assume the framework of the wormhole
calculus and third quantization here, although we use the term baby
universe to describe the disconnected universes. For related work on
baby universes from black hole interiors -- specifically, dynamical
mechanisms by which black hole formation might lead to new universe
creation -- see \cite{baby}. For further analysis of evolution of pure
to mixed states, see \cite{related}, and for some early discussions of
spacetime topology change in string theory, see \cite{strings}.

\begin{figure}[h]
\includegraphics[width=8cm]{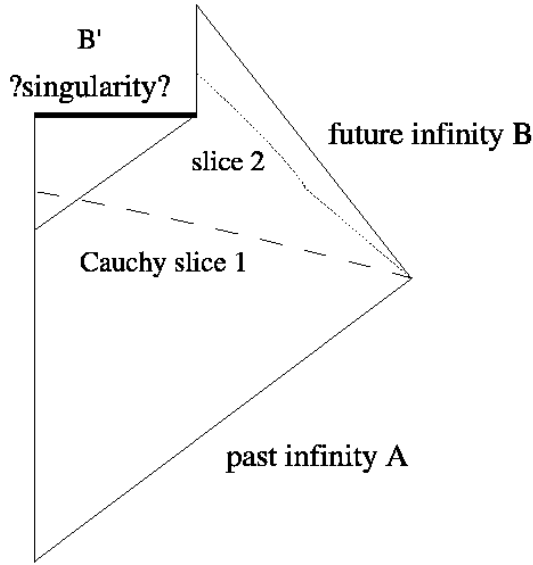}
\caption{Penrose diagram of black hole evaporation. Two spacelike
slices are indicated. Slice 1 is a Cauchy surface, while slice 2 is
not Cauchy surface for the entire universe (parent plus baby)
if black hole formation leads to topology change as in Figure 2.}
\label{penrose}
\end{figure}

\begin{figure}[h]
\includegraphics[width=8cm]{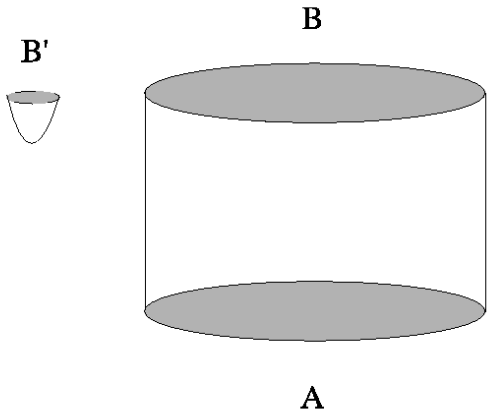}
\caption{Topology change due to black hole formation. The creation of
the baby universe proceeds via quantum gravitational effects and may
lead to a rich internal structure. After
formation, a complete Cauchy surface for the entire universe (combined
spacetimes) must include a slice $B'$ of the baby universe.}
\label{bb}
\end{figure}

\section{Topology change}

Below we describe the specific assumptions of our scenario, which concern
the dynamics quantum gravity (QG), but do not modify physics at large
distances or low energies.

Gravitational collapse leads to a region of Planckian densities and
curvature, where QG effects (fluctuations of the metric) are
large. The size of this QG-dominated region increases with the size of
the black hole, and it likely resolves the singular collapse endpoint
found in classical relativity. It seems plausible, and we assume, that
QG tunneling in this high curvature region can lead to topology change
and a new disconnected spacetime $B'$ which we refer to as a baby
universe. We imagine that it occurs when a large region of high
curvature is formed deep inside the horizon. The tunneling conserves
all charges, and is entirely invisible to observers outside the
horizon -- they see the usual Hawking evaporation of the hole. An
observer falling into the hole hits the QG region after a finite
proper time and her constituents tunnel into the baby universe.  Banks
\cite{review2} has emphasized that the classical geometry of the black
hole interior resembles that of a wormhole with a long throat. We
assume that QG effects cause this throat to pinch off into a baby
universe which is disconnected from the spacetime of the parent
universe. That is, there is no smooth, semiclassical path which
connects the interior of the parent to the interior of the baby
universe.

Any locally conserved quantities such as mass, angular momentum or
gauge charge that would prevent tunneling are still manifest in the
original connected spacetime as black hole hair, and can eventually be
radiated away. Because the Bondi mass (as measured at asymptotic
distance) of the hole does not change, the universe that pinches off
must have zero total energy, and similar arguments apply to its gauge
charge and angular momentum. However, because of negative
gravitational binding energy, this does not preclude a rich internal
structure in the baby universe. One example of how a complex internal
structure might arise is the creation of an inflationary universe from
a finite vacuum bubble \cite{lab}. Indeed, in \cite{lab} the vacuum
bubble appears to exterior observers as a black hole, and the creation
of the sub-universe is caused by quantum tunneling.

A complete specification of the state of the (now topologically
nontrivial) universe requires the wavefunction on $B \cup
B'$. Occupants of $B$, without access to $B'$, have incomplete
information about the universe as a whole, but time evolution $A
\rightarrow B \cup B'$ is completely unitary. (See further discussion
in next section.)

This scenario is similar to that of a remnant, except that in place of
the remnant there is a disconnected region of spacetime which contains
the information that crossed the horizon. We discuss the relation
between the two scenarios below.

It is possible, but not necessary, that information return via QG
tunneling after some long timescale (e.g., exponential in the black
hole mass or area). The energy limitations of the type placed on black
hole evaporation do not apply, as the tunneling modes can have very
long wavelength. If enough (space)time is available, an unlimited
amount of information can be emitted in arbitrarily long wavelength
modes. The returned information emerges from the vertical segment
connected to $B'$ in Figure 1, and eventually reaches future null
infinity $B$. In this case evolution from $A$ to $B$ is completely
unitary, and $B'$ ceases to exist. This case is similar to that of a
decaying, but invisible, remnant.

\section{Evolution from pure to mixed states} 

Objection I, given above, relates to the evolution of pure to mixed
states. Here we explain that this is only a problem if the initial and
final surfaces are both complete Cauchy surfaces.

In quantum field theory, any subset of a Cauchy surface is generically
described by a mixed state, even if the entire Cauchy surface is in a
pure state. It is therefore not surprising if the Hawking radiation
which remains after black hole evaporation is in a mixed state, since
all of it crosses the incomplete slice 2 in Figure 1.

Black hole formation and evaporation from an initial pure state can be
consistent with ordinary quantum mechanics if the final state extends
beyond future infinity in some way. In our case the information
resides in the topologically disconnected baby universe.

Banks, Peskin and Susskind \cite{BPS} identified problems for {\it
local} dynamics (i.e., generalizations of Schrodinger evolution) which
evolve a pure state into a mixed state. In the topology change
scenario, there is no local dynamics which causes evolution from pure
to mixed states. Mixed states only enter the description if one
discards (traces over) some degrees of freedom of the final state,
i.e., by considering only a subset of a final Cauchy surface.

For example, consider two entangled particles, initially in a pure
state. Suppose one of the two falls into the hole and ends up in the
baby universe. The wavefunction describing the two particles evolves
unitarily and is still a pure state even after the hole evaporates,
although the Hilbert space describing the entire universe (parent plus
baby) is then of the form $H = H_{\rm baby} \otimes H_{\rm
parent}$. The state of the two particles is at all times a vector in
$H$, and not a density matrix. A mixed state is only obtained if, in
order to obtain a description of the system in terms of degrees of
freedom remaining in the parent universe, one traces over those
degrees of freedom which fall past the horizon (i.e., we trace over
$H_{\rm baby}$ to obtain a density matrix valued only on $H_{\rm
parent}$).

Two examples illustrate why there is no problem. In both cases {\it it
is the incompleteness of the final region of description -- not
dynamical evolution -- that leads to a mixed state and no pathological
phenomena, such as energy non-conservation \cite{BPS}, are expected.}

1) Wald hyperboloid (Figure 3). Wald \cite{Wa} gave the example of
   evolution a massless scalar field from a complete Cauchy surface
   (e.g., any spacelike slice in flat spacetime) to a surface such as
   a hyperboloid which is entirely contained in the interior of a
   future lightcone. The state of the scalar field on the hyperboloid
   will be mixed even if it was originally pure on the initial Cauchy
   surface, and evolution is unitary. This is simply due to the
   incompleteness of the hyperboloid as a Cauchy surface. If
   correlations exist between modes which intersect the hyperboloid
   and those which do not (because they travel on the lightcone), then
   tracing over these lightcone modes leaves a mixed state on the
   hyperboloid.

2) Inflation (Figure 4). In this example two entangled qubits
   experience an inflationary expansion which leaves them many horizon
   distances apart by the end of the expansion. A local description of
   either qubit requires tracing over the other qubit degrees of
   freedom, leading to a mixed state even if evolution is unitary. An
   observer in a causal patch near one qubit will see the other qubit
   exponentially redshifted as it reaches the de Sitter horizon, so
   that access to its quantum information is clearly lost. The horizon
   of an individual qubit is of course not a complete Cauchy surface.

\begin{figure}[h]
\includegraphics[width=8cm]{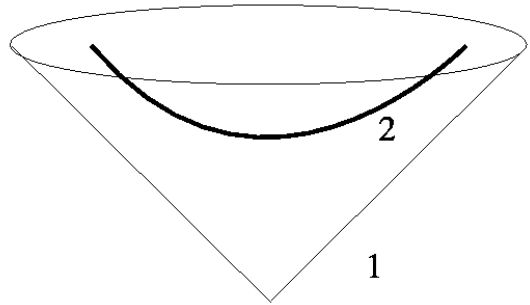}
\caption{Wald's example: Surface 2 is a hyperboloid contained within
the interior of the light cone. Evolution from surface 1 to 2 produces
a mixed state because modes traveling on the lightcone do not
intersect 2.}
\label{hyper}
\end{figure}

\begin{figure}[h]
\includegraphics[width=8cm]{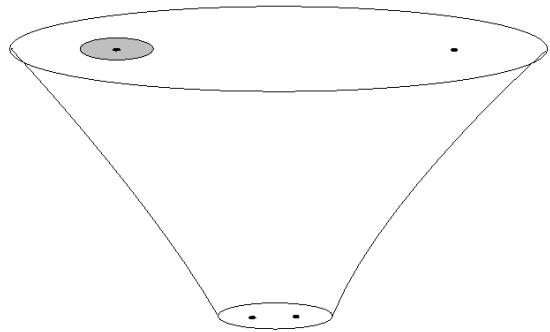}
\caption{An epoch of inflation can cause two initially entangled
qubits (represented by dots) to become separated by exponentially many
horizon distances. A local description of the region near either qubit
will be a mixed state, even if the original state was pure and the
evolution unitary.}
\label{inf-fig}
\end{figure}

\section{Cluster decomposition}

Topology change by black holes can lead to violation of cluster
decomposition, which is objection II above. This point was emphasized
in \cite{Su}, and led Polchinski and Strominger \cite{PS} to propose a
third quantization scenario for baby universes, with consequent loss
of predictability similar to that due to spacetime wormholes. Here, we
note that the violations of cluster decomposition are likely to be
unobservably small, even in hypothetical gedanken
experiments. (Although, as stressed by Susskind \cite{Su}, they remain
vexing as a question of principle in the definition of probabilities
for outcomes of processes involving black hole evaporation.)

Baby universes are distinguishable only by their internal state. By
translation invariance (or general covariance) they carry no memory of
the coordinates of the black hole collapse that led to their
creation. Consider two black hole collapses at spacelike separated
points $x_1$ and $x_2$. {\it If} the internal states $\vert B_1
\rangle$ and $\vert B_2 \rangle$ of baby universes $B_1$ and $B_2$ are
identical (or at least have nonzero overlap), the additional
``crossed'' amplitude (see Figure 5) 
where $B_2$ is created by the collapse at $x_1$
and $B_1$ created by the collapse at $x_2$ must be included, which
violates cluster decomposition (i.e., independence of spacelike
separated events). To obtain the density matrix describing the state
of the parent universe after both black holes have evaporated, one
traces over lost degrees of freedom. This trace receives crossed
contributions if the overlap $\langle B_1 \vert B_2 \rangle$ is
nonzero, and can violate clustering.

\begin{figure}[h]
\includegraphics[width=8cm]{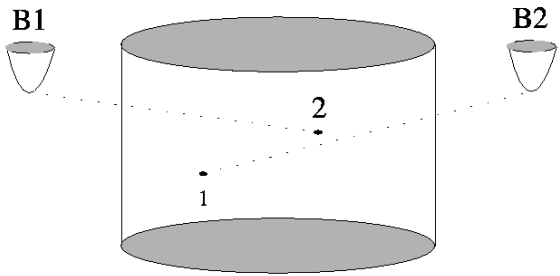}
\caption{A baby universe is only characterized by its internal state.
$B_1$ and $B_2$ may have been created by the black holes at coordinate 1 and
2, respectively, or vice versa. The ``crossed'' amplitude is shown.}
\label{inflation}
\end{figure}

However, it is essentially impossible to realize such interference
processes with macroscopic black holes. The interference is suppressed
if the overlap of internal states $\vert B_1 \rangle$ and $\vert B_2
\rangle$ is small. For example, the charges, masses and angular
momenta of the initial states from which the black holes form must be
equal to within semiclassical uncertainties, or else the Hilbert
spaces describing the two interiors will be different. Even if the two
internal Hilbert spaces coincide, they are necessarily of very high
dimension for a macroscopic collapse (i.e., dimension equal to the
number of degrees of freedom, for example Avogadro's number). As the
dimensionality of a space becomes large, the inner product (overlap)
of any two normalized vectors chosen at random goes to zero -- i.e.,
most vectors are nearly orthogonal. This means that even a small
deviation between the initial configurations which form $B_1$ and
$B_2$ will suppress the interference effect.

An experimenter trying to realize the effect does not, for example,
have control over vacuum fluctuations near or within the horizon of
each hole, and consequently cannot produce identical internal states,
even if the initial collapsing matter configurations are
identical. The simplest way to see this is to note that even if the
initial configurations are identical the observed Hawking radiation
from each hole will be slightly different, which means the internal
states of $B_{1,2}$ are not the same. We expect small perturbations of
otherwise identical initial conditions to produce very different
internal states $\vert B_{1,2} \rangle$, because the infalling matter
experiences strong dynamics at Planck densities.  Although the average
(semiclassical) properties of $\vert B_{1,2} \rangle$ may be similar,
because of the large number of degrees of freedom the overlap $\langle
B_1 \vert B_2 \rangle$ will be extremely small, and hence any
interference effect unobservable.

As a concrete example of the mechanism described above, we can imagine
that interactions are strong enough that the infalling matter has
thermalized by the time it reaches Planck densities and pinches off to
form a baby universe. Then, the internal state of the baby universe
will be a particular Hilbert space vector on a constant energy surface
(microcanonical ensemble). Such vectors have similar coarse grained or
statistical properties (average energy, particle distribution, etc.),
but any two chosen at random will be nearly orthogonal: their overlap
squared decreases as the number of dimensions of the constant energy
surface, which is very large. Under strong dynamics, two almost
identical initial matter configurations will evolve into very
different (essentially, random) state vectors on the constant energy
surface, so it is impossible (highly improbable) to choose initial
conditions that guarantee a large overlap between the internal states
$\vert B_{1,2} \rangle$.

The suppression of cluster decomposition violation discussed above
applies to baby universes with large numbers of internal degrees of
freedom. Presumably, the area result for black hole entropy precludes
small black holes (i.e., with masses of order the Planck energy, and
therefore only a few bits of entropy) from producing complex baby
universes, so topology change on Planck scales would still lead to
cluster decomposition problems. However, it may be that only large
holes produce baby universes, and those always have rich internal
structure, for example due to a mechanism analogous to that in
\cite{lab}. Smaller black holes would then behave as long lived, but
unstable, remnants, without the usual problems (see next section)
since there are only a finite number of species. Such remnants
ultimately return all information back to the parent universe.

For example, the probability of baby universe formation could be
strongly dependent on the black hole size, varying from close to zero
for small holes to unity for large semiclassical holes. In this hybrid
scenario any virtual effects \cite{BPS} due to information loss from
the parent universe would be suppressed by a factor exponential in
$M_*$, the mass scale at which a black hole can be considered
semiclassical (e.g., a large number times the Planck mass).

\section{Relation to remnants} 

There are obvious parallels between our topology change scenario and
that of remnants \cite{review2}. The potentially enormous amount of
information stored in a remnant instead disappears into a baby
universe. In the remnant case one imagines that the throat of the
distorted spacetime region connecting the black hole horizon to the
classical singularity is somehow stabilized, rather than pinching off.

The main problem with remnant scenarios is that there must be a
distinct, long-lived, roughly Planck mass species of remnant for each
black hole which could possibly be formed. With such a large number of
species, it is hard to see why virtual processes involving remnants
would be suppressed -- the multiplicity factor is enough to overcome
even an exponentially small coupling. This concern is alleviated in
our case, as the baby universes do not manifest themselves directly in
the parent universe, and any long lived remnants arising from small
black holes are finite in number.

In the remnant language our proposal can be summarized as follows,
with $M_*$ some scale an order or magnitude or so larger than the
Planck mass. (1) small black holes $(M < M_*$) lead to somewhat long
lived (but not eternal) remnants, whose evaporation is unitary. (2)
large black holes ($M > M_*)$ evaporate as well, but internally cause
topology change and loss of information from the parent
universe. However, the consequences of this apparent non-unitarity
are, for the reasons discussed, small and not excluded by
experiment. Any virtual effects of large black holes are suppressed
exponentially in $M_*$ (as $e^{-M_*^2}$, for example). If
information somehow tunnels back from the baby universes to the parent
universe, this tunneling is manifested in ordinary (but long
wavelength) degrees of freedom, and there are only a finite number of
remnant species in our scenario.

%\bigskip
\section{Conclusions}

We discussed a solution of the black hole information paradox
which depends entirely on details of Planckian physics -- no
modifications of low-energy physics, such as non-locality, are
required.

The main assumptions are that the interior evolution of large black holes produces topology change, and that the quantum gravitational
dynamics of pinching off are strongly coupled. Thus, small
perturbations to the initial state of a black hole lead to different
internal state vectors describing the resulting baby universe, even if
the semiclassical properties are only slightly changed. Under this
assumption, any violation of cluster decomposition will be practically
unobservable.

If this scenario is correct, there is no violation of causality or
locality at the semiclassical black hole horizon, and no stable Planck
mass remnant of black hole evaporation. Instead, much as Hawking first
proposed, information is lost: to a baby universe, from which it may
or may not someday emerge via tunneling. If the information emerges
again, evolution within the parent universe is unitary. If information
remains in the baby universe, the parent universe appears to evolve
from a pure to mixed state, but the evolution of parent {\it and} baby
together is unitary. There are no dire consequences, such as energy
non-conservation.

%\begin{acknowledgments}
\section{acknowledgements}

The author thanks T. Jacobson, B. Murray, J. Preskill and
A. Strominger for useful comments, and S. Giddings in particular for
clarification of the third quantization approach. T. Jacobson provided
a number of useful references to earlier work. This research supported
by the Department of Energy under DE-FG02-96ER40969.

%\end{acknowledgments}

%%%%%%%%%%%%%%%%%%%%%%%%%%%%%%%%%%%%%%%%%%%%%%%%%%%%%%%%%%%%%%%%%
%%%
%%%                     BIBLIOGRAPHY
%%%
%%%%%%%%%%%%%%%%%%%%%%%%%%%%%%%%%%%%%%%%%%%%%%%%%%%%%%%%%%%%%%%%%

\bigskip

%\newpage
%\vskip .75 in


\baselineskip=1.6pt
\begin{thebibliography}{99}



\bibitem{Hawking} S.W. Hawking, Commun. Math. Phys. {\bf 43}, 199
(1975); Phys. Rev. D {\bf 14}, 2460 (1976).

\bibitem{review1}  A. Strominger, Les Houches lectures, hep-th/9501071. 

\bibitem{review2} T. Banks, hep-th/9412131. 

\bibitem{review2a} S. B. Giddings, hep-th/9412138.

\bibitem{review2b} T.~Jacobson, gr-qc/9908031.

\bibitem{review3} L. Susskind and J. Lindesay, The Holographic
Universe, World Scientific, Singapore, 2005.

\bibitem{Gi} S. B. Giddings and M. Lippert, Phys. Rev. D {\bf 69}, 124019
(2004). S. B. Giddings, hep-th/0606146, hep-th/0604072, hep-th/0605196.

\bibitem{BPS} T. Banks, M. Peskin and L. Susskind, Nucl. Phys. {\bf
B244}, 135 (1984).

\bibitem{Su} L. Susskind, hep-th/9405103. 

\bibitem{Dyson} F. Dyson, IAS preprint (1976), unplublished.

\bibitem{St} A. Strominger, hep-th/9405094; see comments on this paper
in \cite{Su} above.

\bibitem{PS} J. Polchinski and A. Strominger, Phys. Rev. D {\bf 50}
7403 (1994), hep-th/9407008.

\bibitem{Ja} T. Jacobson, gr-qc/0308048.

\bibitem{EB}  D.~A.~Easson and R.~H.~Brandenberger,
   JHEP {\bf 0106}, 024 (2001), hep-th/0103019.

\bibitem{AB}
 A.~Ashtekar and M.~Bojowald,
  Class.\ Quant.\ Grav.\  {\bf 22}, 3349 (2005), gr-qc/0504029;
  Class.\ Quant.\ Grav.\  {\bf 23}, 391 (2006), gr-qc/0509075.
 
\bibitem{wormholes} S. Coleman, Nucl. Phys. B {\bf 307}, 864 (1988);
S. B. Giddings and A. Strominger, Nucl. Phys. B {\bf 307}, 854 (1988).

\bibitem{baby}
 V.~P.~Frolov, M.~A.~Markov and V.~F.~Mukhanov,  Phys.\ Lett.\ B {\bf 216}, 272 (1989), Phys.\ Rev.\ D {\bf 41}, 383 (1990); 
C.~G.~Callan, S.~B.~Giddings, J.~A.~Harvey and A.~Strominger,
  Phys.\ Rev.\ D {\bf 45}, 1005 (1992), hep-th/9111056;
  V.~F.~Mukhanov and R.~H.~Brandenberger,  Phys.\ Rev.\ Lett.\  {\bf 68}, 1969 (1992);
  R.~H.~Brandenberger, V.~F.~Mukhanov and A.~Sornborger,
  Phys.\ Rev.\ D {\bf 48}, 1629 (1993), gr-qc/9303001;
  M.~Trodden, V.~F.~Mukhanov and R.~H.~Brandenberger,
  Phys.\ Lett.\ B {\bf 316}, 483 (1993), hep-th/9305111;
  A.~A.~Tseytlin and C.~Vafa,
  Nucl.\ Phys.\ B {\bf 372}, 443 (1992), hep-th/9109048];
  G.~L.~Alberghi, D.~A.~Lowe and M.~Trodden,
  JHEP {\bf 9907}, 020 (1999), hep-th/9906047;
L.~Smolin,  Class.\ Quant.\ Grav.\  {\bf 9}, 173 (1992);  gr-qc/9503027;
 astro-ph/9712189.

\bibitem{related} M. Srednicki, Nucl.\ Phys.\ B {\bf 410}, 143 (1993),
 hep-th/9206056; S. B. Giddings, Phys.\ Rev.\ D {\bf 49}, 4078 (1994), 
hep-th/9310101; W.~G.~Unruh and R.~M.~Wald, Phys.\ Rev.\ D {\bf 52}, 2176 (1995), hep-th/9503024.

\bibitem{strings} P.~S.~Aspinwall, B.~R.~Greene and D.~R.~Morrison,
Phys.\ Lett.\ B {\bf 303}, 249 (1993), hep-th/9301043, and Nucl.\
Phys.\ B {\bf 416}, 414 (1994), hep-th/9309097; E.~Kiritsis and
C.~Kounnas, Phys.\ Lett.\ B {\bf 331}, 51 (1994), hep-th/9404092.

\bibitem{lab} S.~K.~Blau, E.~I.~Guendelman and A.~H.~Guth, Phys.\
Rev.\ D {\bf 35}, 1747 (1987); E.~Farhi and A.~H.~Guth, Phys.\ Lett.\
B {\bf 183}, 149 (1987); E.~Farhi, A.~H.~Guth and J.~Guven, Nucl.\
Phys.\ B {\bf 339}, 417 (1990); A.~D.~Linde, Nucl.\ Phys.\ B {\bf
372}, 421 (1992).
 
\bibitem{Wa} R. M. Wald, Quantum Field Theory in Curved Spacetime and
Black Hole Thermodynamics, University of Chicago Press, (1994).



%\end{references}
\end{thebibliography}
\end{document}